\documentclass[9pt,sigconf]{acmart} 
\renewcommand\footnotetextcopyrightpermission[1]{} 

\usepackage{booktabs} 
\usepackage[T1]{fontenc}
\usepackage[utf8]{inputenc}
\usepackage{balance}
\usepackage[printonlyused]{acronym}
\usepackage{xcolor}
\usepackage{comment}
\usepackage{amsmath}
\usepackage[font=small, labelfont=bf, belowskip=-10pt, aboveskip=5pt]{caption}

\acrodef{FEC}[FEC] {Forward Error Correction}
\acrodef{ARQ}[ARQ]{Automatic Repeat Request}
\acrodef{QoS}[QoS]{Quality of Service}
\acrodef{QS}[QS]{Qualitative Service}
\acrodef{BPP}[BPP]{Big Packet Protocol}
\acrodef{IP}[IP]{Internet Protocol}
\acrodef{NDN}[NDN]{Named Data Networking}
\acrodef{ICN}[ICN]{Information-Centric Networking}
\acrodef{TSN}[TSN]{Time-Sensitive Networking}
\acrodef{SLO}[SLO]{Service Level Objectives}
\acrodef{ARVR}[AR/VR]{Augmented Reality/Virtual Reality}
\acrodef{NPU}[NPU]{Network Processing Units}
\acrodef{ECN}[ECN]{Early Congestion Notification}

\AtBeginDocument{%
  \providecommand\BibTeX{{%
    \normalfont B\kern-0.5em{\scshape i\kern-0.25em b}\kern-0.8em\TeX}}}
    
\setcopyright{acmcopyright} 

\begin{document}
\title{A Framework for Qualitative Communications Using Big Packet Protocol}

\author{Richard Li$^1$, Kiran Makhijani$^1$, Hamed Yousefi$^1$, Cedric Westphal$^{1,2}$, Lijun Dong$^1$} \author{Tim Wauters$^3$, Filip De Turck$^3$}

\affiliation{\vspace{0.3cm}
  \institution{$^1$ Futurewei Technologies, Santa Clara, CA, USA\\$^2$ Computer Engineering Department, University of California, Santa Cruz, CA, USA\\$^3$ IDLab, Ghent University -- imec, Ghent, Belgium}
}
\email{{richard.li, kiranm, hamed.yousefi, cedric.westphal, lijun.dong}@futurewei.com, {tim.wauters,filip.deturck}@ugent.be}

\renewcommand{\shortauthors}{R. Li et al.}

\begin{abstract}
In the current Internet architecture, a packet is a minimal or fundamental unit upon which different actions such as classification, forwarding, or discarding are performed by the network nodes. When faced with constrained or poor network conditions, a packet is subjected to undesirable drops and re-transmissions, resulting in unpredictable delays and subsequent traffic overheads in the network. Alternately, we introduce qualitative communication services which allow partial, yet timely, delivery of a packet instead of dropping it entirely. These services allow breaking down packet payloads into smaller units (called chunks), enabling much finer granularity of bandwidth utilization. 

We propose \texttt{Packet\,Wash} as a new operation in forwarding nodes to support qualitative services. Upon packet error or network congestion, the forwarding node selectively removes some chunk(s) from the payload based on the relationship among the chunks or the individual significance level of each chunk. We also present a qualitative communication framework as well as a \texttt{Packet\,Wash} directive implemented in a newly evolved data plane technology, called Big Packet Protocol (BPP).
\end{abstract}

\copyrightyear{2019} 
\acmYear{2019} 
\acmConference[NEAT'19]{ACM SIGCOMM 2019 Workshop on Networking for Emerging Applications and Technologies }{August 19, 2019}{Beijing, China}
\acmBooktitle{ACM SIGCOMM 2019 Workshop on Networking for Emerging Applications and Technologies (NEAT'19), August 19, 2019, Beijing, China}
\acmPrice{15.00}
\acmDOI{10.1145/3341558.3342201}
\acmISBN{978-1-4503-6876-6/19/08}

%
%



\maketitle

\section{Introduction} 
Transport control methods such as flow and congestion control are responsible for reliable and in-order delivery along with ensuring the integrity of the received information. Any error, due to link congestion or intermittent packet loss in the network, can trigger re-transmission of data packets. This results in unpredictable delays as well as an increase in the network load, wasting network resources/capacity. To mitigate this problem, different schemes have been proposed such as in data centers~\cite{WANG,DCLAT,PCUT}, media streaming~\cite{RFCAQM, FEC:1}, and wireless networks~\cite{FLEXCAST}. While some of these schemes are based on mechanisms for efficient and faster re-transmissions, and others utilize redundant transmissions, we propose a novel approach that attempts to eliminate or at least effectively reduce the re-transmissions in the network. This is critically important especially in emerging applications, such as holographic telepresence and tactile Internet, which require extremely low latency and high throughput.	

For all packet-based network architectures such as \ac{IP} or \ac{ICN}~\cite{Zhang14,ccnx,Ghasemi18,Ghasemi1818}, a packet is a minimal, self-contained unit of delivery that gets transmitted, classified, or discarded indiscriminately by the network nodes. Packet size is expected to increase with the increased support for jumbo frames~\cite{jumbo}. We contend that the network should not drop the entire packet upon packet error or network congestion, but rather break it down into smaller logical units with each unit (called chunk) having its own significance-factor describing its importance in the context of information carried in the payload (or having a relationship with other chunks). In contrast to current major service models in networking (i.e., differentiated services \cite{DIFFSERV} and integrated services \cite{INTSERV}), we are here dealing with the subjective quality of the packet itself, i.e., what aspects of a packet are relatively more significant than others. As the quality associated with chunks may vary from each other, such services in networks may be referred to as {\em qualitative services}. 

We provide the network with an opportunity to manipulate and modify the packets by introducing \texttt{Packet\,Wash} as a new technique and an alternate approach to packet drops. It allows partial, yet timely, delivery of packets by removing less significant chunks from them as needed. While original host's applications take care of packetization, forwarding nodes are in charge of \texttt{Packet\,Wash} operation. We 
present several possible techniques to utilize \texttt{Packet\,Wash} operation. We propose a framework to identify a qualitative context as well as qualitative packet format. We finally employ a newly evolved data plane technology, called Big Packet Protocol (BPP), to implement \texttt{Packet\,Wash} for qualitative services. BPP attaches meta-information or directives into packets, guiding intermediate routers on how to process the packets. 	

The rest of this paper is organized as follows. Section~\ref{what} introduces qualitative communications. The \texttt{Packet\,Wash} technique is proposed in Section~\ref{Qtechniques}. Section~\ref{framework} discusses a generic framework required to build qualitative services. Section~\ref{sec:BPP} demonstrates a realization of \texttt{Packet\,Wash} in the data plane. Section~\ref{related} presents the related work. Section~\ref{discussion} discusses the limitations and extensions of the proposed approach before concluding with Section~\ref{concl}.
\section{Qualitative Service Concept}
\label{what}

The \ac{QoS} functions ensure that packets marked with higher priority are scheduled earlier than those with lower or normal priorities. As a consequence, under adverse network conditions such as resource congestion, the lower priority packets get completely dropped. Re-transmission of packets can waste network resources, reduce the overall throughput, and cause both longer and unpredictable delays in packet delivery. Not only the re-transmitted packet has to travel part of the routing path twice, but the sender does not realize the packet has been dropped until timeout or negative-acknowledgment happens, which also adds to the extended waiting time at the sender side before re-transmission is initiated. The current approach of handling the packet error or network congestion, discarding the packet entirely, is expensive.

We contend it would be beneficial if a forwarder could selectively drop parts of the packet payload to reduce the packet size and alleviate congestion while forwarding the remainder of the packet to its destination. This is where the qualitative communication service comes into play. It provides a packetization method that breaks down the payload into multiple chunks, each with a certain significance. The network nodes understand the significance or relationship of the chunks and accordingly make decisions to drop chunk(s) based on the current situation, such as congestion level, the priority carried in the packet, etc. The significance associated with chunks may vary from each other. 

Chunks with higher significance are less likely to be dropped when qualitative services are applied in the network. As an example for video streaming, the sender could rearrange the bits in the payload such that the first consecutive chunks contain the base layer, while the next chunks the enhancement layers. Thus, in case of congestion, a forwarding node can intentionally remove the chunks containing enhancement layers as many as necessary.

Optionally, the chunks in the packet payload may have a certain relationship among each other. For example, we can use a network coding scheme where the chunks actually sent are linearly coded from the original chunks in the payload and are linearly independent from each other. Thus, dropping any of the linearly coded chunks would still keep the rest of chunks useful to the receiver to recover the original data contained in the packet payload (see~\ref{encoding}). 

The qualitative service is a native feature of the networks, which has the following benefits: (a) a packet re-transmission may not be needed if the receiver has the capability to comprehend what is left in the packet after removal of certain chunks from the payload by the intermediate network nodes, and the receiver can recover as much information as needed. In this case, the receiver can acknowledge the acceptance of the packet, while it may also indicate to the sender that it was partially dropped in the network. Network resource usage can be tremendously reduced and better prioritized for the delivery of other packets; and (b) the latency of packet delivery can be significantly reduced due to the absence of re-transmissions. Some of the information contained in the original packet can be recovered by the receiving node, as long as some recovery algorithms or methods are agreed and known in advance by the sender, the forwarding nodes, and the receiver. The algorithms and methods can be carried along with the packet, such that it can be detected and executed by the intermediate network nodes, and revealed to the receiver, which can carry out the reverse operation to recover some or all the information contained in the packet.

\section{Qualitative communication techniques}
\label{Qtechniques}

In this section, we introduce three techniques to realize qualitative communications in the network: (1) \texttt{Packet\,Wash}, 
(2) adaptive rate control, 
and (3) in-packet network coding. Although all techniques are dealing with \texttt{Packet\,Wash} operations in the forwarding nodes, their main focuses are on in-network operations, feedback control in transport layer, and resiliency in applications, respectively. 

\subsection{Generic Packet Wash} 
\label{packetwash}
In qualitative communications, \texttt{Packet\,Wash} can be seen as a scrubbing operation that reduces the size of a packet while retaining as much information as possible. 
It operates by dropping lower-priority chunks from the payload according to the information carried in the packet header, helping the forwarder to understand the significance of (or the relationship between) the chunks. The lost chunks may not be recovered but some information is usable at the receiver.

\begin{figure} []
\includegraphics[width=0.7\columnwidth]{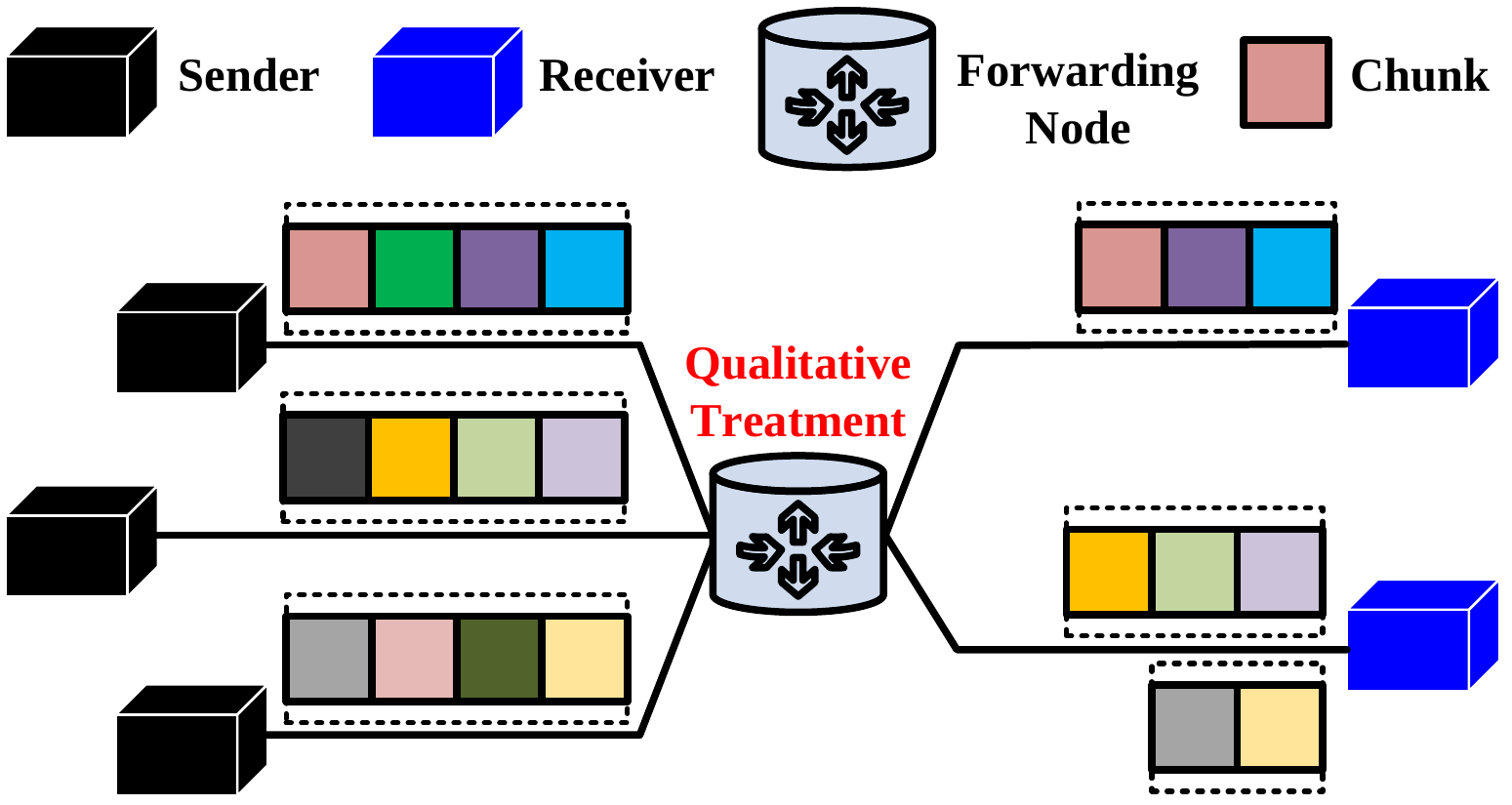}
	\caption{\texttt{Packet\,wash} drops some chunks in intermediate nodes}
	\label{fig:Qtrimming}
\end{figure}

Note that creating the qualitative packet is not part of \texttt{Packet\,Wash} and is accomplished at the sender nodes. \texttt{Packet\,Wash} is thus only an operation in the intermediate forwarding nodes. 
Fig.~\ref{fig:Qtrimming} shows how a forwarding node treats the payloads of three incoming packets in case of congestion while providing partial, yet useful, content delivery.

\textbf{Qualitative Entropy}---\texttt{Packet\,Wash} uses meta-data in the header, denoted as qualitative entropy (q-entropy), to alter the payload. Thus, a qualitative packet is required to specify: (1) a function through which network nodes treat a packet; (2) a chunk-dependent significance parameter understood by this function; (3) the threshold beyond which a packet cannot be further degraded (as it would become useless); and (4) the network condition when it is to be treated. This collectively defines q-entropy. A \textit{washed packet} is a result of the q-entropy being applied, i.e., for any packet \(p\), if washed packet is \(p'\) and \(Q_f \) is the q-entropy, then \(p'\) = \(Q_f(p)\). The operation \(Q_f(.)\) can be applied until the washed packet reaches the threshold at which it cannot be further degraded. $p' = \lim_x Q_f^x(p)$, where $x$ represents the number of operations at successive forwarding nodes. The function is applied when a particular condition is met, and the degradation threshold \(T\) has not been reached yet. If that happens, the payload cannot be further reduced because it will be rendered unusable.  

\subsection{Adaptive Rate Control}

As part of \texttt{Packet\,Wash} operation, chunks may be dropped from the packet. \texttt{Packet\,Wash} performs selective trimming of a payload from less to higher significant chunks. Accordingly, the forwarding node makes a decision on which packet to trim and for this packet, which chunk(s) to trim. Until the network conditions improve, the receivers receive poor quality streams. This may become undesirable over a period of time. 
In order to ease the network load and reduce congestion, the receiver can check the modified header of a \textit{washed packet} and trigger an adaptive congestion-control by notifying the sender about the level of congestion in the network as per Section~\ref{transport}. It can send an acknowledgment with a quality of packet value, which the sender uses to alter its transmission rate. This adaptive rate control utilizes network resources more effectively. Moreover, it significantly reduces data delivery delay by partially delivering the packets as well as dynamically managing packet sizes, which is critically important in emerging real-time applications.
 
This use of quality of packet improves network efficiency and fairness among users. In particular, if a packet that has been qualitatively treated already is in contention for buffer space with a packet that has not been trimmed yet (all other priorities being equal), the forwarding element should trim the intact packet.
Moreover, since a packet that has had its payload reduced is more congestion-friendly, the forwarding layer should give it a higher priority. A simple scheme would be to increase the QoS level (i.e., degree of significance) for priority level chunks that are being dropped, while regular traffic and full packets would be served as Best Effort.

It is worth mentioning that the trimming operations are not restricted to only tail-drops as a chunk may be removed from anywhere within the payload since this provides higher flexibility for applications to categorize significance. However, the forwarding nodes will lower overheads when only tail drops are done because then the amount of buffer shift is minimized. Therefore, a particular trimming approach chosen by applications will need to consider the trade-off between performance and flexibility.

\subsection{In-Packet Network Coding}
\label{encoding}
Network coding~\cite{ho2006random,ahlswede2000network} is a networking technique in which the data is encoded, transmitted, and then decoded in order to increase the resiliency of the network. Network coding has been widely used at the packet level~\cite{sundararajan2009network}, where multiple packets are encoded and transported between the same source and destination. In case of congestion, an entire packet is dropped, which means the receiver would lose one degree of freedom to decode the whole coded packets, and the packet delivery would take a longer time.

In this section, we propose to apply random linear network coding to the chunks in the packet payload. In other words, we increase the network-coding granularity from packets to chunks. The data to be transmitted is segmented into groups of $k$ chunks. These $k$ chunks are then network-coded together into $k' \geq k$ new chunks. From these network-coded chunks, we create payloads by inserting $h$ network-coded chunks into the payload of the packet (where $h$ is the largest number of chunks that fits in a packet). If $k \leq h$, then the whole group of $k$ chunks fits within one packet. On the other side, the receiver needs $k$ chunks (or degrees of freedom) to decode the original data. The qualitative packet header carries the coefficients for the linear combination of each network-coded chunk. 

\begin{figure} []
\includegraphics[width=0.55\columnwidth]{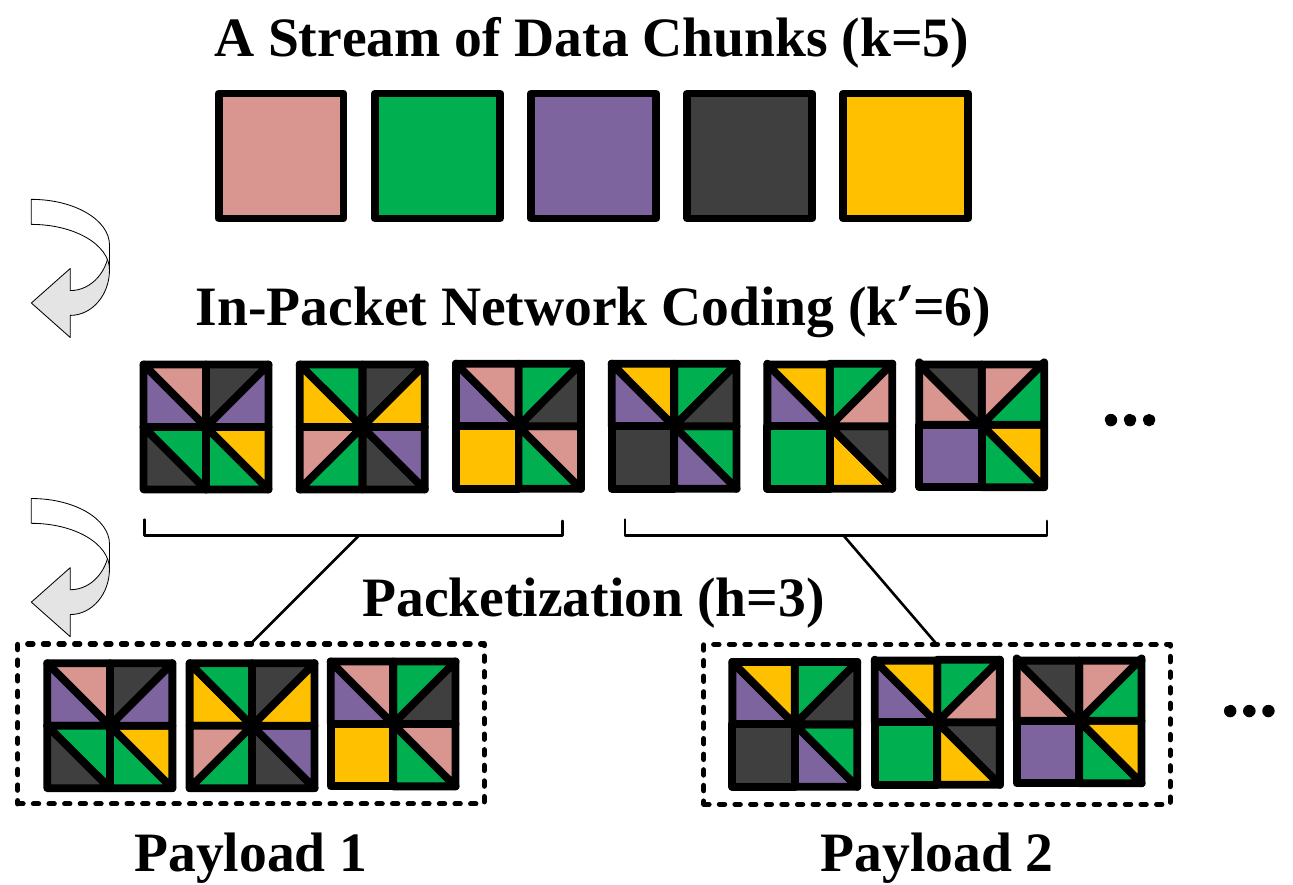}
	\caption{In-packet network coding}
	\label{fig:networkcoding}
	\vspace{-.2cm}

\end{figure}

The proposed approach is illustrated using a simple example in Fig.~\ref{fig:networkcoding}. 
The receiver can acknowledge the number of degrees of freedoms it has received, and the sender can keep generating packets with new combinations until $k$ degrees of freedom have been received to decode the original data. The sender does not need to know which chunks are lost along the way. It only needs to send more (linearly independent) chunks of newly coded packets, in a number equal to the missing degrees of freedom.

Qualitative communication services can be facilitated by utilizing \texttt{Packet\,Wash} on coded packets. The sender transmits the packet with coded chunks, which may be partially dropped by intermediate forwarding nodes. The benefit of applying random linear network coding to the chunks of the packet payload lies in the following aspects:

\begin{itemize}
\vspace{-0.05cm}
\item The sender can add some ratio of redundancy in the packet payloads, as shown in Fig.~\ref{fig:networkcoding}. There are five chunks in the group, six coded chunks are added into two packets' payloads with 20\% of redundancy. If any intermediate forwarding node drops any one chunk from the two packets, the receiver can still decode the original packets.

\item When the packet eventually reaches the receiver, any chunks that are retained in the packet can be cached by the receiver and are useful for future decoding of the original payload after enough degrees of freedom are received.

\item When network congestion happens, the intermediate router does not need to decide which chunk to drop, it can randomly select as many chunks as needed until the outgoing buffer permits to contain the packet. There is no need for priority in this context and not need to track which specific chunk has been lost.
\end{itemize}






\section{Qualitative Communications Framework}
\label{framework}

In order to realize qualitative services with one of the above mentioned \texttt{Packet\,Wash} techniques in the network, a modular framework needs to be in place. The framework consists of application, transport, and network components. Firstly, a qualitative context needs to be identified by the applications, discussed in Section~\ref{significance}. Secondly, the transport layer mechanisms for congestion management/control should be modified to support this type of services (Section~\ref{transport}), and finally the packet format to support this capability needs to be defined (Section~\ref{pformat}). 

\subsection{Qualitative Context}
\label{significance}

The characterization of what information is qualitatively more significant is decided by the applications. They also need to describe the number of chunks and their boundaries in the payload. The context is needed to be able to tell what can be removed. A qualitative context includes a function selected by the application which can be applied over the entire payload. The context also associates a significance-factor to each chunk in the payload. The applications on sender nodes provide this context to the network stack. On the receiver side, the context allows them to measure the degraded value and hence derive the quality of the received packet.

An application may utilize the following type of functions in the context:
\begin{itemize}
\item Degree of importance: It can characterize the priority (see Section \ref{pformat}).
\item Recovery function: It can recover or regenerate chunks of the packet using a relationship with the other part of the payload or even across the data flow.
\item Degree of relative significance: A context may indicate relative and statistical significance, i.e., given context, a packet may be derived from earlier delivered packets of a flow.
\end{itemize}

The complete framework needs to support an interface between the application and network layer because only applications can decide how to break the payload into chunks and assign them significance, but this has to be done within a format that is supported within the network. This, in turn, builds the qualitative packet. The details are not presented here due to the space limit. 

\begin{figure} []
\includegraphics[width=0.95\columnwidth]{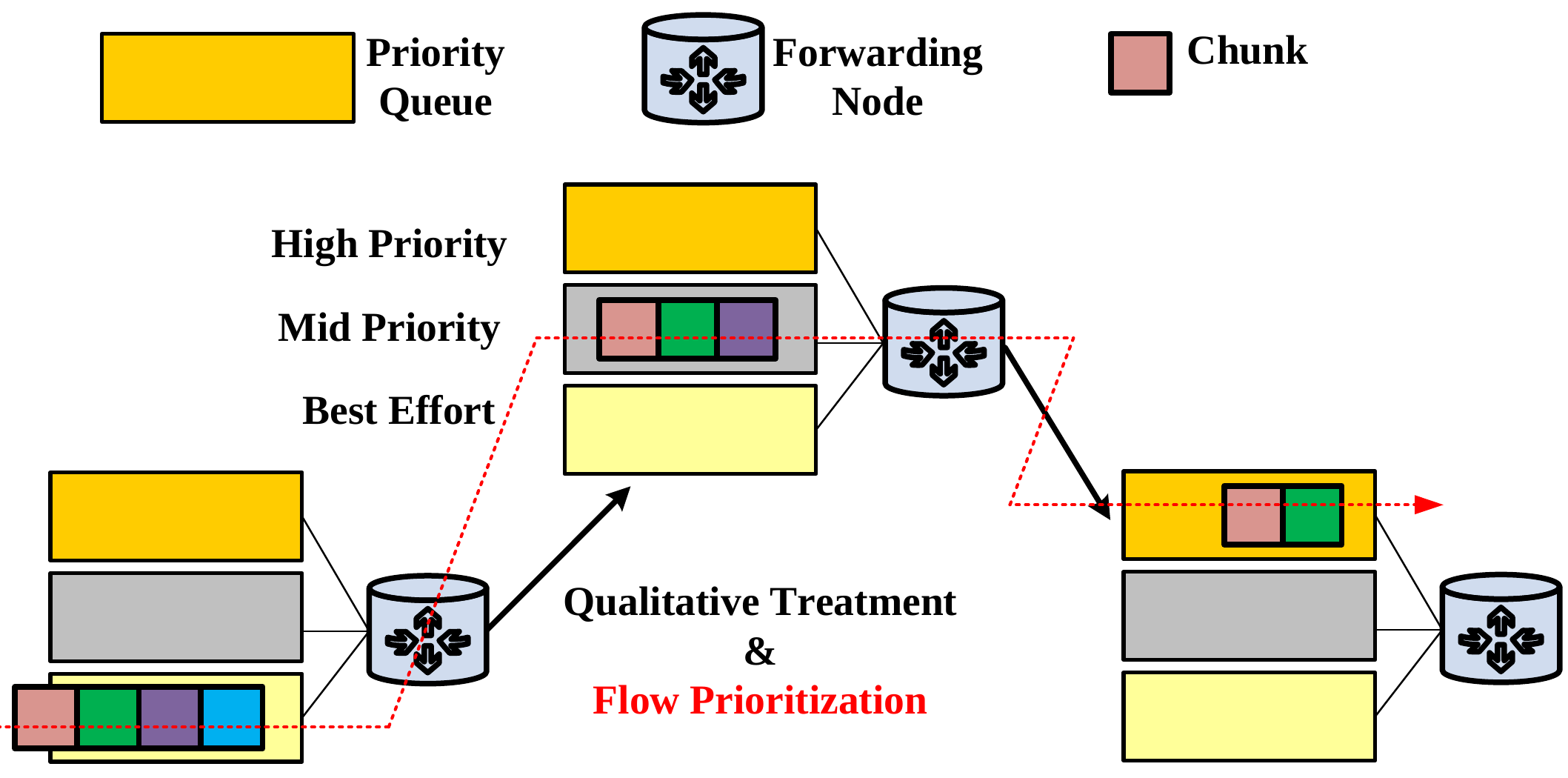}
	\caption{In-network transport prioritization}
	\label{fig:transport}
\end{figure}

\subsection{Transport Layer for Qualitative Services}\label{transport}

A qualitative stream of packets adapts itself to some extent to congestion (by immediately reducing its rate via dropping some chunk(s)). As a consequence, the mechanism to manage congestion should be modified accordingly. In particular, the coexistence of qualitative communication services along with regular, legacy services should be addressed. In an IP network, this would entail modifying part of the transport header. 

Qualitative services require some end-to-end congestion management if only to be compatible with legacy networks. The partial dropping of a packet should be understood by the end-to-end congestion management as a form of \ac{ECN}, that is a warning that some level of congestion is occurring. However, there needs to be a hop-by-hop congestion control as well to decide what chunks to drop from a packet. 

At each forwarding element making a decision to drop chunks from the qualitative packets, it is important to preserve fairness among different flows. One mechanism to achieve this is to increase the \ac{QoS} level of the packet after a \texttt{Packet\,Wash} treatment, by increasing the ToS to the next higher value. Fig.~\ref{fig:transport} shows how the priority of a qualitative packet changes while forwarding through the network nodes. 

Congestion sharing can also be utilized to spread the impact of congestion over a larger number of flows. Because the space in the buffer is shared among more packets (instead of dropping a packet and accepting the next one, both packets may see their payload cut in half), this is more fair in the short term. This has the added benefit of notifying more flows of the congestion occurring inside the network.

\begin{figure} []
\includegraphics[width=0.90\columnwidth]{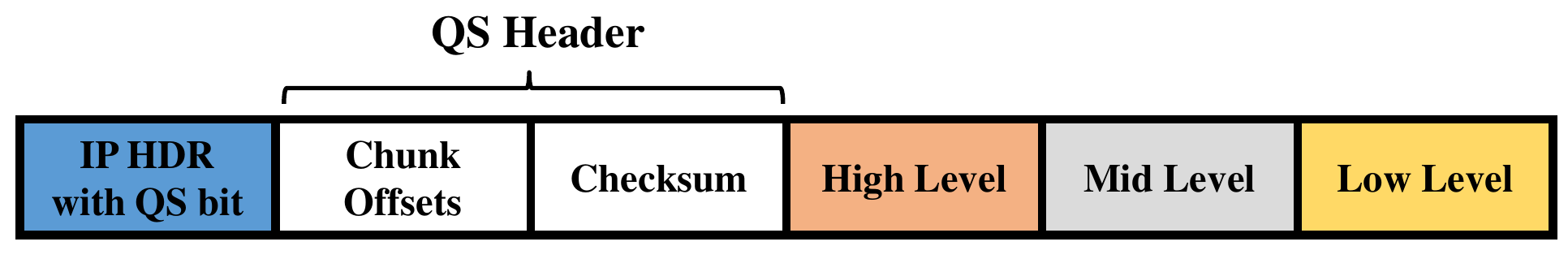}
	\caption{Idealized qualitative packet format}
	\label{fig:pformat}
\end{figure}

\subsection{Packet Format}\label{pformat}

While there are many possible mechanisms to support qualitative communication services, the basic scheme is to insert in the IP header (namely \ac{QS} header) some indication that the packet supports qualitative services. The \ac{QS} header is used to identify the payload structure in form of the logical chunks and their significance-factors. 
For example, as shown in Fig.~\ref{fig:pformat}, this header can indicate what the different priority levels of each chunk in the payload are, and how to identify the payload chunks associated with these priority levels. Using this format, one may consider three priority levels, Gold, Silver, and Bronze, and the \ac{QS} header would indicate to drop Bronze first, then Silver, etc. The number of priority levels would depend on the network and application. For instance, in the data center, it is sometimes beneficial to cut the payload and only forward the header~\cite{PCUT, DCLAT}. This is due to the use of shallow buffers in order to speed up communications. In networks where buffers would fill up more slowly, more priority levels can be supported. 

The QS header could specify a specific offset for each chunk, or refer to a known vector of offsets. A tail-drop policy would be easily implemented if priority levels are ordered accordingly. 

The QS header may include some checksum or CRCs for different chunks so that the integrity of the packet can still be verified even after \texttt{Packet\,Wash} operation. However, packet-level checks should be disabled and replaced by chunk-level checks. 

The QS header can also include a Significance Function (detailed in Section \ref{significance}) that assigns a different significance to each chunk. This could be implicit (as in the Gold/Silver/Bronze example, where the significance is embedded in one of three levels) or explicit. The service delivery will be of tolerable quality if less significant content was lost or delayed, while most valuable content was delivered on-time.


\section{In-network Qualitative Packet Processing}\label{sec:BPP}

Expressing the quality of a packet for easy processing by forwarding engines on network nodes is the first step towards validating the idea. In this section, we use a specific data plane technology, called \ac{BPP}, to describe a generic \texttt{Packet\,Wash} (Section~\ref{packetwash}) processing on forwarding nodes.

\subsection{BPP Overview}
\begin{figure} 
	\includegraphics[width=0.85\columnwidth]{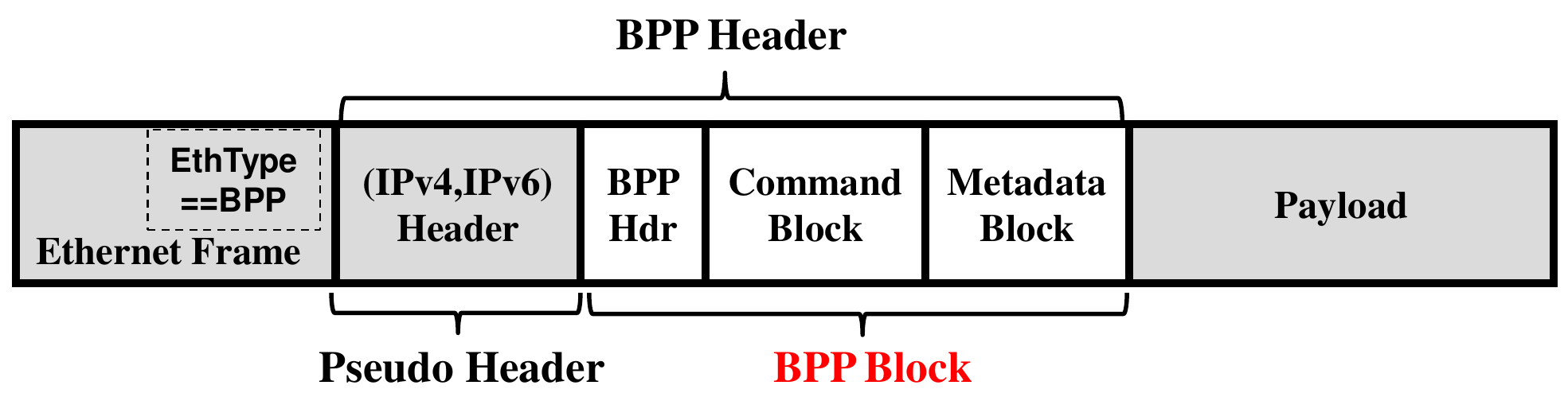}
	\caption{BPP packet format with BPP block}
	\label{fig:format}
	\vspace{0.5cm}
\end{figure}

Big Packet Protocol framework~\cite{BPP} is a programmable data plane technology compatible with IP networks. The basic idea in \ac{BPP} is to attach meta-information or directives using BPP blocks into packets. This meta-information provides guidance to intermediate routers about processing those packets. The so-called BPP block, shown in Fig.~\ref{fig:format}, is in fact a contract between the application and network. 

In a \ac{BPP}-aware network, routers process BPP directives and take corresponding actions. Having a structured syntax, they can be processed by \ac{NPU} within a bounded processing time (or with minimal overhead). 
The BPP framework allows per-packet behavior for functionality such as in-band per-packet signaling facilities, per-flow path selections, and network-level operator decisions. Note that the changes in packet processing operations can affect network equipment~\cite{BPP}. 


Designed for high-precision services~\cite{NET2030}, a BPP contract carries commands for in-time or on-time packet delivery specification. This allows network nodes to schedule in-transit latency accurately. In IP networks, BPP has been applied to \ac{TSN} bridges over large scale networks~\cite{TSNBPP}. It shows that BPP directives help reduce overall configuration overheads for TSN-schedulers. BPP has been extended to different application domains, such as using meta-data for collaborative vehicular information exchange~\cite{COLLAB}, latency guarantees in multimedia streaming~\cite{MMEDIA}, semantic mashup in Internet of Things (IoT)~\cite{MASHUP}, and computation offloading in Mobile Edge Cloud (MEC)~\cite{ComputeOffloading}.

Because of the flexible and structured contract, BPP is inherently suitable for implementing qualitative services. A qualitative packet can be represented by a \ac{BPP} contract consisting of \texttt{Packet\,Wash} directive which will have significance-factors as its meta-data. By doing so, network nodes remain unaware of the user-payload and wash packets only as prescribed by the application.  

\subsection{Packet Wash directive in BPP}
We propose to extend the BPP contract to include a new qualitative declarative, also called \texttt{Packet\,Wash}, the details of which are described below.

\subsubsection{Design considerations}
The design of the \texttt{Packet\,Wash} declarative dictates the behavior of packet with respect to quality and takes the following into consideration:

\begin{itemize}
\item  \textbf{conditional-directive}: Not all BPP-commands are conditional, but \texttt{Packet\,Wash} must be a conditional declarative. It is applied only after determining that the network state is adverse and the outcome is likely that the packet will not reach the receiver.

\item \textbf{q-entropy function}: Each packet carries this function through which the network nodes understand how to operate on the payload based on the significance-factors associated with chunks.

\item \textbf{resource-resolution}: When packets from two or more distinct flows contend for the same resource, with all else being equal, qualitatively treated packets could be given a higher priority.

\item \textbf{latency-constraint}: If a qualitative packet is determined to arrive late at the destination even after qualitative treatment or at the cost of processing, then it is worth to drop it. 
\end{itemize}

\subsubsection{BPP Packet Wash structure}
\begin{figure} 
 	\includegraphics[width=\columnwidth]{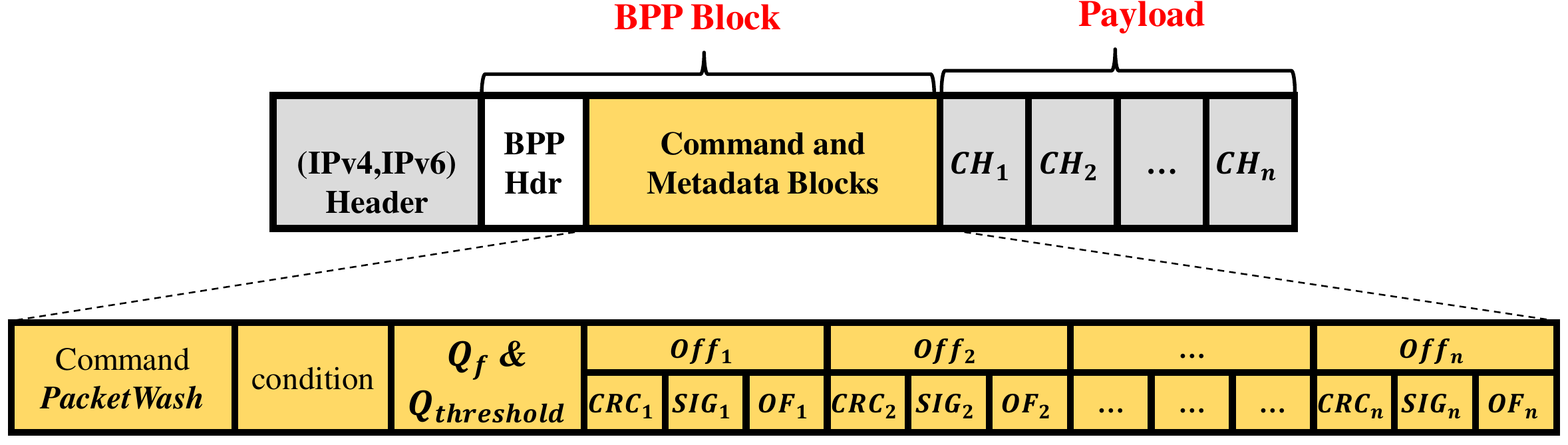}
 	\caption{Qualitative packet using BPP: Packet format and \texttt{Packet\,Wash} directive details}
 	\label{fig:packetwash2}
 \end{figure}

On the basis of our design goals, a representation of the qualitative packet format is shown in Fig.~\ref{fig:packetwash2}, where the payload (i.e., user data) is broken down into chunks, and \texttt{Packet\,Wash} directive is carried as part of the packet header. 

The new format contains the following parameters: (1) a \textit{command} ``PacketWash''; (2) a \textit{condition} when \texttt{Packet\,Wash} is applied; (3) $Q_f$, a q-entropy function that defines an operation on the payload; e.g., priority, binary, or step function; (4) $Q_{threshold}$, a threshold value, called qualitative-threshold, beyond which the chunks cannot be further dropped; and (4) some extra information about each chunk $i$: (a) $SIG_i$, a significance-factor associated with the chunk as per the function; e.g., priority order, or binary 0 or 1 bit; 
(b) \textit{Off$_i$}, an offset to describe the location of the chunk in the payload; (c) $CRC_i$, a CRC to verify the integrity of the chunk; and (d) $OF_i$, a flag to determine if the chunk was dropped. This helps receivers know which chunks have been dropped in the network.

\subsubsection{Processing on forwarding nodes}
On a forwarding node, packet processing is performed in the following steps:

\begin{enumerate}
\item As the qualitative packet arrives, BPP engine extracts \texttt{Packet\,Wash} command and checks for the condition, such as if egress queue is 90\% full.
\item If the condition is true then the engine applies the function in q-entropy to parameters of each chunk in the payload. For example, if a function is binary, parameters have value 0 or 1. The output gives the chunk offset(s) to be dropped.
\item BPP engine drops those many bytes from the offsets of resulting chunks and mark them dropped in the header.
\item BPP engine drops the packet if the degraded quality exceeds the qualitative-threshold, or the packet is still determined to arrive late at the destination; otherwise, BPP engine forwards it to the next node.
\end{enumerate}
\section {Related Work}\label{related}

Qualitative communication services are related to a few existing concepts such as removal of payload to create network bandwidth, notifying the senders to adapt data rates, and minimizing retransmissions. 
A direct relation may be seen with Flexcast~\cite{FLEXCAST}, which works by a confidence estimate of a decoded bitstream to gracefully adapt to video degradation. In contrast to Flexcast, our qualitative service concept focuses on network mechanisms in fixed or wired networks. Moreover, \texttt{Packet\,Wash} is a novel ``significance-based''  scheme and works at the packet level, not at bit-level encoding.

With respect to the trimming method, specific to data centers, \cite{PCUT} and \cite{DCLAT}, are receiver-driven traffic control mechanisms that use packet trimming. In particular, the goal is to achieve fast re-transmissions; therefore, nodes have very shallow buffers. The downside of this scheme is a high risk of packet drops when congestion occurs. In our qualitative \texttt{Packet Wash} trimming technique, the payload is not entirely dropped, nor does it necessarily trigger a re-transmission.

A mechanism to achieve high throughput in video streaming using dynamic fragmentation scheme is proposed in~\cite{IPFRAG}. It groups consecutive MPEG-2 TS (transport stream) packets with same priority into a single IP packet. When network condition deteriorates, large IP packets are fragmented into smaller packets, which results in increasing the probability of reception. Dynamic fragmentation improves the throughput at the cost of duplicating IP headers per fragment. Using our proposed approach, even better goodput can be achieved.
\section{Discussion and Future Work}\label{discussion}

\textit{\textbf{Overhead---}}Qualitative services come with the cost of adding some header in the packet to assist the network with \texttt{Packet\,Wash} operation. A question arises as to the trade-off in terms of bandwidth saved versus the added size of this new header. We argue that this overhead will be minimal in future networks where the size of the payload will increase dramatically, and the size of the MTU as well. As a quick rule of thumb, we can assume that the \ac{QS} header will differentiate between $p$ chunks (potentially associated with a priority level) and that each chunk will require an additional 16 bits check-sum. The header itself should be able to point out to the offsets in the payload that start each chunk. This can be accomplished with 11 bits per offset for an MTU of 1280 bytes, as in IPv6. It can also be achieved with fewer bits if the forwarding nodes have some priory information regarding where the offsets could be. For instance, a few well-known possible payload formats could be provisioned ahead of time and listed in the qualitative services header. This means that $4p$ bytes may suffice to describe the different chunks or priority layers. For a handful of priority levels, it means that the overhead of qualitative services is minimal, in the order of 1\% for three levels. With larger MTU (jumbo frames can carry 9,000 bytes payload), the overhead is even less.

However, the gain may be significant. As a simple illustrative use case, we can consider an immersive video streaming~\cite{he2018network,westphal2017challenges,VID360} application. The sender may transmit the whole 360-degree view to the receiver but may assign a lower priority to views that are far from the current Field of View (FoV) of the receiver. In this case, the use of qualitative services may reduce the bandwidth by roughly 5/6th, that is over 80\%. While this is a very simple example, it points towards significant benefits for qualitative services. 

Besides, some computational overhead may be imposed to the forwarding nodes (e.g., to recompute/update the offsets after dropping some chunks), which is negligible.

\vspace{0.5cm}
\textit{\textbf{Encryption and Tunneling Considerations---}}There are several host level functions that get applied to the payload. The mechanisms to implement encryption varies in qualitative services since now the payload received is a subset of the original. However, this does not mean that the integrity of the packet cannot be preserved. A suggested approach is to encrypt chunks independently. Similarly, other host-originated payload specific functions can also be applied on each chunk separately. Note that this is neither optimal nor more secure, therefore, this is an area for future research.

Another concern that may arise is how to handle/treat qualitative packets while repackaging the traffic data in forwarding nodes for tunneling. This way an entire actual qualitative packet is encapsulated into the payload of a new packet that hides the nature of the traffic running through a tunnel, thus challenging the functionality of qualitative communications. One straight-forward solution is to extract the qualitative-related headers from the actual packets to the tunneled ones. However, an optimal solution would be part of our future work.

As a future study, we intend to analyze the results of simulation of qualitative services under different network conditions. The paper presented a discrete method of using the chunks for identifying quality. Possibility of other mathematical models that may be applied to the payload without chunking it are yet to be explored.
\section{Conclusion}\label{concl}

We introduced qualitative communications as a new service to the Internet, enabling much finer granularity of bandwidth utilization. We also introduced \texttt{packet\,wash} as a novel network operation which manipulates context-aware packets in intermediate nodes based on the subjective quality of the packet itself, i.e., what aspects of a packet are relatively more significant than others. We proposed a framework to realize qualitative communications, and finally described a possible implementation of a \texttt{Packet\,Wash} directive using Big Packet Protocol (BPP).

Qualitative services guarantee the delivery of critical information. We believe that qualitative communications make a paradigm shift from today's QoS mechanisms. This concept will play a key role in realizing a new way of delivering a multitude of services; for example, ``qualitative'' rendering of a web content, guaranteed delivery of critical messages, and next-generation of multimedia applications that consume high-bandwidth and offer real-time experiences such as Augmented Reality (AR) and holographic media.

\balance
\bibliographystyle{ACM-Reference-Format}
\bibliography{qualitative-bibliography}

\end{document}